# Complementing cell taxonomies with a multicellular functional analysis of tissues

Ricardo Omar Ramirez Flores[1,*], Philipp Sven Lars Schäfer[1], Leonie Küchenhoff[1], and Julio Saez-Rodriguez[1,*]

[1] Heidelberg University, Faculty of Medicine, and Heidelberg University Hospital, Institute for Computational Biomedicine, Heidelberg, Germany

* Corresponding authors:
JSR: pub.saez@uni-heidelberg.de; + 49 6221 56 7521
RORF: roramirezf@uni-heidelberg.de; + 49 6221 56 7521

**Abstract**

The application of single-cell molecular profiling coupled with spatial technologies has enabled charting cellular heterogeneity in reference tissues and in disease. This new wave of molecular data has highlighted the expected diversity of single-cell dynamics upon shared external queues and spatial organizations. However, little is known about the relationship between single cell heterogeneity and the emergence and maintenance of robust multicellular processes in developed tissues and its role in (patho)physiology. Here, we present emerging computational modeling strategies that use increasingly available large-scale cross-condition single cell and spatial datasets, to study multicellular organization in tissues and complement cell taxonomies. This perspective should enable us to better understand how cells within tissues collectively process information and adapt synchronized responses in disease contexts and to bridge the gap between structural changes and functions in tissues.

**Introduction**

Cooperation among the diverse cells in tissues is essential for the proper functioning of the majority of physiological processes. Multicellular coordination relies on mechanisms that couple the intracellular processes of individual cells (eg. metabolism, signaling and gene regulation) with extracellular processes with other cells and abiotic components, like the extracellular matrix. Studies of the immune (1), cardiovascular (2), and endocrine system (3), among others, have shed light on the mechanisms regulating multicellular coordination. For example, in the control of cardiovascular function, the endothelium works as a network of organized collaborative sensors that in parallel process multiple signals from the microenvironment relying on information sharing across neighboring endothelial cells (4). To properly address and coordinate stable responses to a constantly changing microenvironment, tissues exploit cellular diversity that enables division of labor across cells or enhances overall efficiency and adaptability (4–14). Thus, to understand tissue function we need to explore not only the variety of cells in the human body and their spatial organization, but also their multicellular coordination.

Initiatives such as the Human Cell Atlas (15), the Human Biomolecular Atlas Program (16), the Kidney Precision Medicine Project (17), and the Human Tumor Atlas Network (18), have committed to the mission of charting the human cellular diversity of distinct organs, locations (here referred to as organ compartment), and physiological contexts. Growing collections of single cell transcriptomic atlases across organs and conditions are confirming the known diversity of genetically identical cells (19, 20) and generating an extensive single cell taxonomy that defines, names, and organizes cell types. This taxonomy groups single cells based on their lineage and/or potential functions using data-driven clustering approaches together with prior knowledge on cell identity markers. These cell-centric taxonomies of tissues have challenged the definitions of what a cell type is (21) and raised discussions on how to define proper cell ontologies (22–25). By generating highly detailed human tissue descriptions across clinical conditions, we hope to learn more about disease processes, which will eventually help us to develop better therapies. But whether this becomes a reality depends on finding the relationships between cellular diversity, multicellular organization, and tissue physiology.

The question remains whether we can use available single-cell, spatial and bulk omics data to study how diverse cells are coordinated in tissues to perform physiological functions. Emerging approaches aim to reframe the analysis of single-cells in a tissue-centric manner (26–30), where the objective is to infer high-level representations of multicellular responses, such as coordinated changes of gene expression across groups of cells upon a stimulus, and later decompose them into specific cell-cell dependency networks. This perspective requires combining the information of the compositions of cell identities in a tissue, their spatial arrangement and the covariance of their molecular features. By shifting the analysis focus from individual cellular behaviors to tissue function, it may be possible to 1) study the coordination between cells, 2) study cellular diversity as a mechanism to ensure collective function, and 3) compare various multicellular processes in human physiology.

In this perspective article we discuss emerging strategies for the analysis of single cell data to study multicellular coordination in tissues and complement the descriptions of cellular diversity. We start by briefly reviewing the current efforts to describe cellular diversity from single cell data and discuss their potential limitations to identify cell groups that impact tissue function. We discuss how by studying cellular coordination it may be possible to assign functional relevance to groups of cells that play specific roles in multicellular processes. Then, we summarize the known basic hallmarks of multicellular coordination in tissues and explore the possibilities of inferring multicellular processes from available single cell data. We finish by discussing the potential utility of the adoption of a multicellular perspective to analyze single cell data that could help us to expand our understanding of the use and regulation of single cell diversity for tissue function.

**The taxonomy of tissues and its functional limit**

Variability among cells is a natural and emerging characteristic of functional relevance within cell populations (10, 31–38). Advances in sequencing and imaging technologies (39, 40) have enabled scientists to study the diversity of single cells in tissues at large scale. Current efforts, particularly in biomedicine, have focused firstly on charting the molecular diversity of single cells from human tissues in different clinical conditions, where the main objective is to classify, name, and organize cells based on their developmental origin (cell-types), functional profile, and/or location. For example, cardiac muscle cells are differentiated from skeletal muscle cells because they have distinct molecular profiles (41); similarly, within cardiac cells, ventricular and atrial cardiac muscle cells can be separated into two distinct classes based on their gene expression (42), and upon heart damage, cardiac cells that upregulate stress genes can be separated from the cardiac muscle cells in homeostasis (43). However, the molecular profiles of individual cells are confounded by technical factors (e.g. RNA capture efficiency, bias, and scale (44)) and difficulties in sampling equivalent regions across patients. Therefore, the single-cell field has prioritized the integration of large numbers of single-cell profiles from independent studies to make more precise descriptions of the molecular diversity of cells in tissues (45).

In general practice, tissues across clinical conditions are mostly compared in terms of the compositions and spatial organizations of these cell classifications. This taxonomic understanding of tissues has benefits in defining recurrent patterns of tissue architecture across different samples and contexts. For example, an emerging group of cells can be identified in certain disease conditions (46–50), or specific cell-type spatial neighborhoods can be associated with tissue states (51–56). The classification of tumors as hot and cold, based on the levels of T-cell infiltration is a canonical example of how the composition of the tumor microenvironment relates to cancer disease progression and therapy response (57). The currently dominating framework of single cell taxonomy often leads researchers to profile tissues to identify diverse classes of cells and their spatial arrangements, assuming that this can provide significant insights into the functional state of the tissue (Figure 1A).

Analysis of cross-condition single-cell transcriptomic atlases has shown that cells with similar molecular phenotypes can be found in different organ compartments, tissues, and clinical contexts (Figure 1B) (46, 58–61). Additionally, the use of spatial technologies to profile human intact tissues are confirming that cells can behave similarly with distinct cellular neighbors or physiological conditions (28, 43, 58). For example, in the heart, cardiomyocytes upregulating stress markers have been identified not only upon ischemic damage, but also close to the vasculature in healthy cardiac tissues (43). Moreover, it has been shown that disruptions of cellular contacts between cardiomyocytes lead to the emergence of their stressed state (62). These observations highlight the importance of considering the cellular context and spatial relationships in understanding and interpreting cell groups and their impact on tissue function across diverse conditions.

By listing all the possible cell phenotypes in tissues using transcriptomics, we aim to explain tissue function. However, the extent to which cell taxonomic categories reflect cellular function remains uncertain. Three issues may contribute here: First, categorizing cells into discrete groups does not accurately represent the continuous nature of biological functions (Figure 1C) (63). Second, cell taxonomies are built using limited information of their phenotype, usually mRNA levels within a cell. For example, in an analysis of single mesenchymal cells, it was demonstrated that the expression of extracellular matrix marker genes did not accurately predict its deposition (64). It may be expected that similar results can be observed in other cell types across tissues and contexts (65), suggesting the existence of non-functional cellular classifications within the current defined taxonomies mainly built from mRNA levels (Figure 1D). Third, cell function is dependent on cooperation. For example, a given cell group may perform a function only in the presence of a specific microenvironment. An illustration of this phenomenon is shown by Shalek and colleagues in an early time course single-cell transcriptomics profiling of dendritic cells upon pathogenic stimulations (66). In their work, they reported the existence of rare precocious cells with high core antiviral activation in early time points upon stimulation. They observed that as time progressed, the number of cells activating the antiviral program increased via paracrine signaling that reinforced the molecular phenotype observed initially only on rare cells. When dendritic cells were studied in isolation, the pathogenic stimulation did not drive an increase in abundance of cells activating an antiviral program, suggesting that rare precocious cells were necessary to coordinate the population antiviral response. In this example, rare dendritic cells could be interpreted as first responders or leaders of a multicellular process associated with pathogenic stimulation (Figure 1E). Other similar examples in the immune system show the importance of rare cells for collective behaviors that impact tissue function (67–69), supporting the need for a multicellular understanding of cell heterogeneity. Given the inherent diversity of cells, a key challenge of the taxonomic approach is to identify cell groups with an impact on tissue function. Incorporating cellular relationships into the taxonomic framework may aid in attributing a functional role to groups of cells (70, 71).

As the number of described cell identities increases with the generation of independent single-cell transcriptomic atlases, it is relevant to explore the dependencies between distinct groups of cells for community-level coordination and its impact on tissue function. What is the link between cellular diversity, single-cell decision making and community function? Given the diverse behavior of single-cells, what are the community-level mechanisms that orchestrate apparently deterministic functions in tissues? Is the deregulation of community processes a hallmark of human disease? All of these questions require extensions to the current analysis frameworks established as best practices for the analysis of single-cell data and strategies tackling some of these questions are starting to emerge as discussed in the following sections.

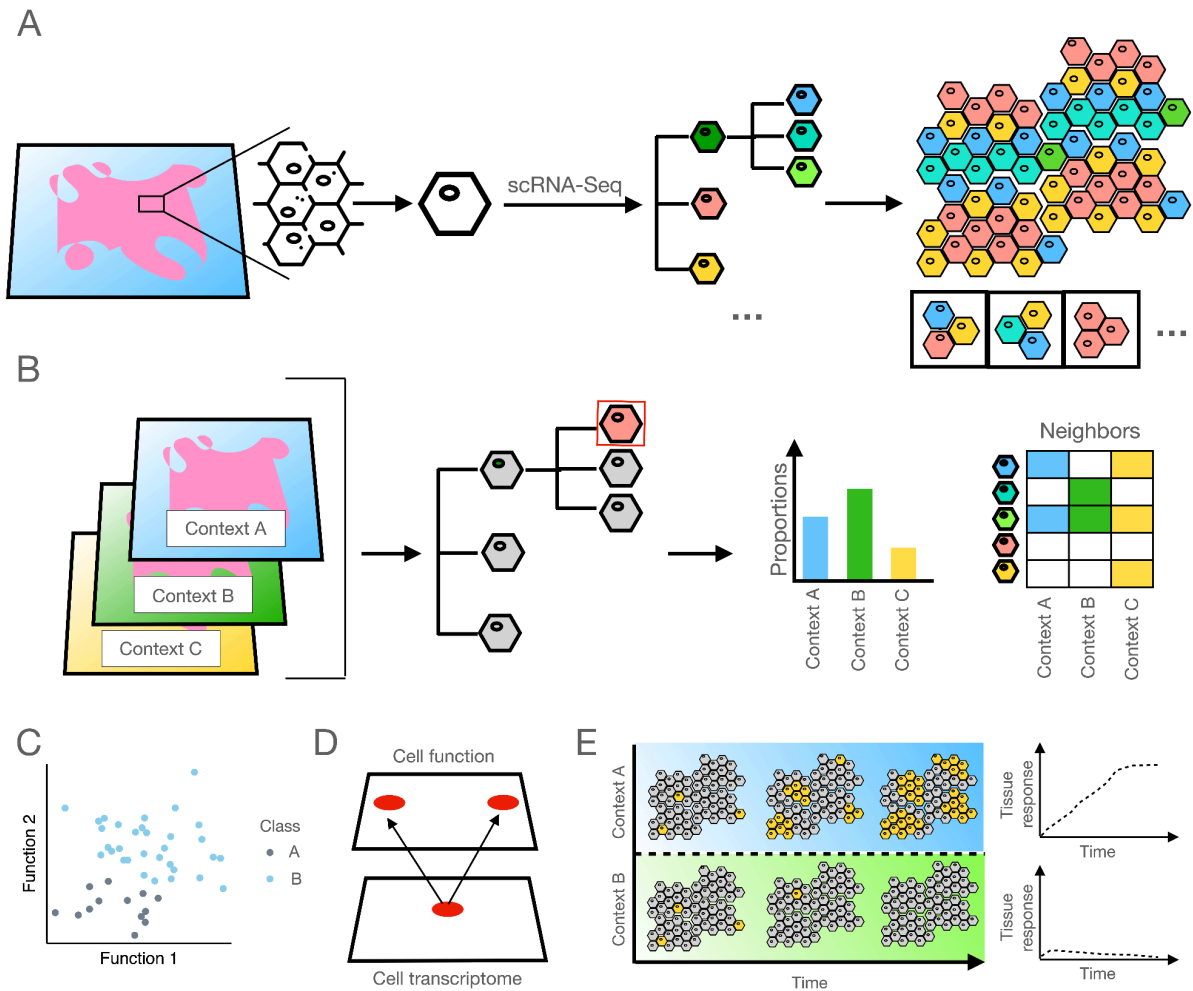

**Figure 1 Single cell taxonomies and their limitations.**

A) Tissue collections undergo molecular profiling at the single-cell level using technologies like single-cell RNA sequencing (scRNA-Seq) that quantify thousands of molecular features simultaneously. Data-driven classifications yield a taxonomy summarizing the variability and functions of single cells within a tissue and allow for the definition of cell identities. In addition, spatial technologies profiling single cells in intact tissues define an organizational taxonomy, reporting the spatial arrangements of specific cell identities.

B) Cross-condition single-cell transcriptomic atlases reveal similar cell phenotypes in various organ compartments and clinical contexts, while spatial technologies profiling intact tissues confirm that cells exhibit comparable molecular profiles with different cellular neighbors or physiological conditions.

C) Classification of cell phenotypes into discrete taxonomic classes does not take into consideration the continuous characteristics of biological functions.

D) Cell taxonomies are built using limited information of their phenotype, usually mRNA levels within a cell, thus it is possible that non-functionally relevant cell groups exist within the current defined taxonomic classes. One example is the apparent discordance between the transcriptome of mesenchymal cells and their capabilities to deposit extracellular matrix.

E) Incorporating cellular relationships into the taxonomic framework may aid in attributing a functional role to groups of cells. For example, rare cell groups may represent first responders or leaders of multicellular processes upon specific tissue contexts. In the schematic we show the state of two tissues throughout time in two distinct processes (e.g. pathogenic and mechanical stimulation). In early time points in both contexts rare cells with similar molecular profiles can be identified (yellow cells). However, as time progresses, only in one of the contexts, these cells increase in composition by reinforcing their phenotype and associate with the activation of a tissue-level process.

**Multicellular information processing in tissues**

From development and organ physiology to aging and disease, multicellular coordination is an essential characteristic of human biology (72). Thus, cellular function cannot be studied without its context and its relation to other cells, and tissue function cannot be understood solely as the sum of independent cellular processes, but rather as an emergent property of their dependencies. The emergence, evolution, and convergence of multicellularity through life history (73) are complex questions that evolutionary biologists are tackling with experimental and theoretical frameworks and are largely unknown. Although the evolutionary perspective of multicellularity is extremely relevant to understand human physiology, for our discussion we will focus on the analysis rather than the origin of multicellular processes in human biology.

In human biology, the study of multicellular coordination has historically focused on cell-cell communication through direct contacts and signaling molecules (74). Extensive work on cellular communication has shown the mechanisms by which cells influence each other's behaviors and generate functional spatial patternings via signaling pathways (75, 76). Experimental work on multicellular systems, such as yeast models emulating diffusing-dependent organs and islets of Langerhans (endocrine cells), has demonstrated the critical role of multicellular dependencies in physiological functioning. For instance, glycolytic synchronization waves in yeast and synchronized insulin production in islets of Langerhans depend on cell communication and multicellular dependencies (77–80). Moreover, in aging, disruptions of stem cell niches can alter the balance between self-renewal and commitment processes (81). While the importance of multicellular activity in physiological processes is evident, the regulatory mechanisms ensuring proper large-scale multicellular responses at the tissue level remain elusive. These mechanisms would need to modulate the variability of single-cell responses depending on the context. This current gap in knowledge is partly due to the absence of highly-multiplexed technologies capable of profiling individual cells in their spatial context through time, and manipulating cellular heterogeneity in multicellular systems. Nonetheless, the combination of single cell imaging analysis, multicellular engineered systems, and time-course perturbational experimental designs has provided insights into the mechanisms of multicellular information processing (14, 77, 82–84).

Regarding the interplay between cellular heterogeneity and multicellular coordination, theoretical studies have proposed that cell communication may impact the diversity of cellular processes within a population (Figure 2A) (14). This relationship depends on the intrinsic variability of intracellular processes, where communication tends to reduce variability in highly stochastic situations and enhance it when intracellular processes are less variable. Drawing parallels with microbial communities, where phenotypic heterogeneity plays a role in quorum sensing systems (85), similar mechanisms may operate in human multicellular processes (11). For example, bet hedging in multicellular systems could offer diverse solutions for rapid responses to changing environments without relying on individual cells. However, the specific mechanisms regulating cellular dependencies for controlling variability in response to sudden microenvironmental changes remain unclear.

Multicellular coordination in human cells has been studied in short term responses (e.g. calcium export) of *in vitro* cell collectives upon cyclic mechanical or chemical perturbations. In these studies, multicellular coordination is followed dynamically to identify the emergence and maintenance of dependencies between individual cells, forming what we term multicellular information networks. In these networks, each node represents a cell, and cellular dependencies, such as cell-cell communication, are depicted as edges. Work in endothelial (82), neuronal (83), and endocrine cells (3) has consistently revealed similar dynamics in these multicellular information networks, transitioning from local to global organization (Figure 2B). Initially, small interconnected networks led by a few "first-responders" or communication hubs emerge in response to a perturbation. Subsequently, groups of cells follow these leaders over time. Long-term responses that follow are characterized by increased dependency across cells, reflected by an increment in communication hubs and a decrease of cellular diversity (82). To facilitate this transition, multicellular information networks rely on memory and reinforcement capabilities (Figure 2B). Memory implies that individual cells' roles in the information network—specifically, the degree of information they transmit or receive—correlate across consecutive stimulus cycles. Reinforcement refers to the gradual strengthening of these correlations over time and stimulus cycles. In addition, robustness of multicellular responses has been shown in pancreatic islet cells, where the removal of first-responders or hub cells lead to the emergence of new ones as long as there is no significant disruption to cellular dependencies (86, 87). Although all of these concepts require further validation in diverse cellular contexts and *in vivo*, studies on the multicellular dynamics of specific physiological processes offer a fundamental understanding of how information networks adapt within tissues.

Further understanding of multicellular coordination in physiological processes faces several challenges. It is required to extend the principles of multicellular coordination by including cells of diverse lineages and long-distance communication events, such as paracrine signaling. Moreover, due to the complexity of simultaneously measuring different cellular processes over time, most of our knowledge of multicellular function is reduced to a single output, such as calcium export. In addition, it is necessary to consider the information carried by mechanical interactions (88), extracellular vesicles (89), and the intercellular space (eg. extracellular matrix) (90), that may be essential in the organization of multicellular responses. From the current research, the spatial rules governing the organization of the different cellular roles in the information networks of (patho)physiological processes are still poorly described. Although cellular interactions are constrained by the physical arrangement of cells in tissues, it is relevant to question if the location of first-responders have certain spatial patterns that ensure the proper transition from a local to global coordination described before. Alternatively, as in other nucleation processes it could be the case that first responders emerge stochastically. Finally, memory and reinforcement have been described in short term and cyclic perturbations of multicellular systems; if these adaptations translate to long-term tissue adaptations, such as cardiac remodeling, fibrosis or sustained immune responses, is not known.

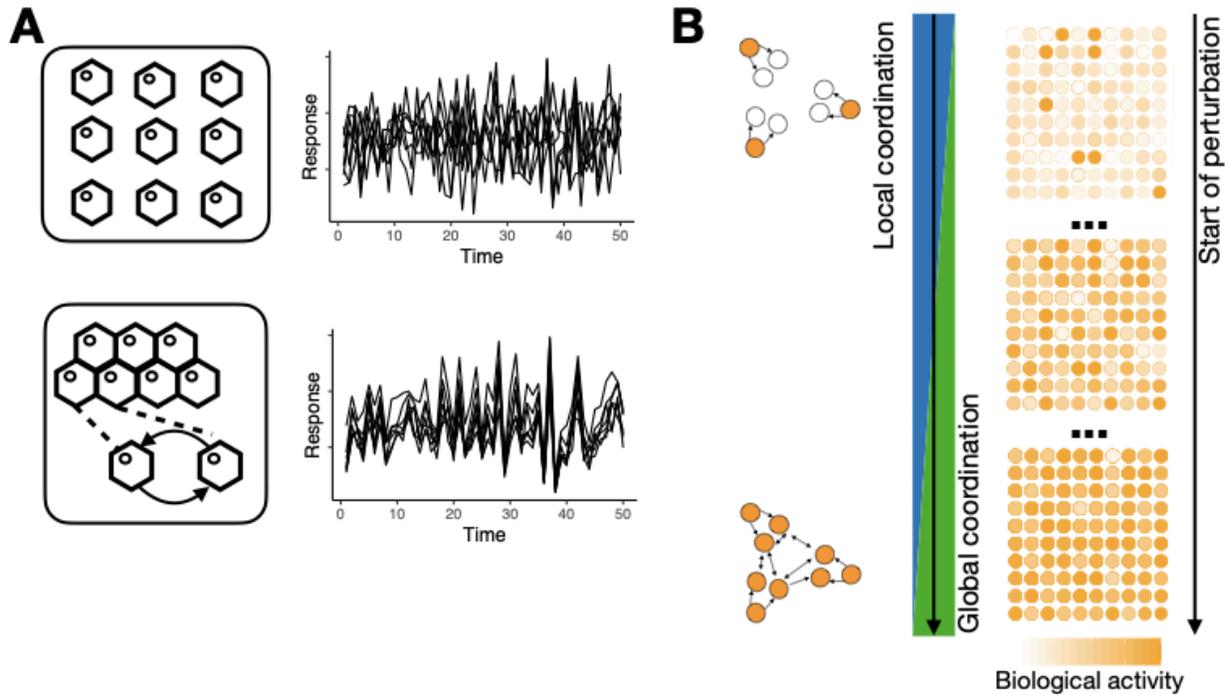

**Figure 2 Principles of multicellular coordination.**

A) Cells in isolation show variable responses to perturbations (upper panel). Cell communication facilitates communal function by regulating the levels of variability (lower panel). In this example we illustrate how cell communication reduces the variability in the cell's response and enables coordination.

B) Work in multicellular systems has shown that multicellular coordination transitions from local responses to global coordination via memory and reinforcement of biological activities.

**Inference of multicellular information networks from single-cell data**

Can we use available single-cell and spatial data to infer and study multicellular information networks that describe (patho)physiological processes? We believe that by combining descriptions of tissue composition and organization with measurements of cellular coordination, it may be possible to study disease processes from the lens of multicellularity.

Specifically, this view would require not only to look at the compositions and spatial organizations of the tissue components (e.g. cell-groups), but also to include the relationship between these building blocks to coordinate large-scale tissue functions. As a consequence, this view focuses on generating descriptions of the state of the tissue, an umbrella term referring to the multicellular function of the tissue, where particular cell-groups are part of a latent coordinated higher-order process(Figure 3A).

With single-cell data from dissociated and intact tissues, compositional and organizational descriptions of tissues at a molecular resolution have increased (Figure 3A). Compositional descriptions are natural consequences of the taxonomic perspective, where tissues are represented as collections of building blocks in different compositions (91–93). These building blocks can be cell-types or functional cell-groups, or, in the case of spatial data, neighborhoods or niches, which are local and repeated patterns of cellular or functional co-occurrence (94–97). Organizational representations of tissues describe potential interactions across cells, usually constrained by spatial distance or signaling relations. In the case of single-cell data, a popular strategy that is used to derive organizational features is cell-cell communication inference. At its core, these methods infer potential signaling between groups of cells, by relating the expression of the ligand of one group with the receptor of another one, limiting interactions based on prior knowledge (98). Given that these methods focus on co-expression patterns at the transcript level (while signaling is largely mediated by proteins), and they are limited by uncontextualized ligand-receptor interactions, they represent an indirect measurement of cell communication (99). In high- and low-resolution spatial -omics data, tissue schematics are built to generate a hierarchy of cellular or functional organization from the local to the global scale (100). These methods rely either on predictive tasks where an observed feature in a location is explained by its neighboring locations (101, 102) or by counting the recurrence of local interactions (103–105). These strategies output dependencies of cells or functions constrained by their location. However, since each spatial interaction is usually interpreted independently they provide a limited description of multicellular coordination in the tissue as a whole.

A collection of novel computational methods has started to tackle multicellular coordination associated with the tissue state (Figure 3B). These multicellular integration methods infer multicellular programs from multi-sample cross-condition single cell or spatial data, starting from a defined tissue taxonomy consisting of cell-types, cell-groups or spatial niches (26–30). A multicellular program in these methods represents co-varying molecular features (e.g. gene expression) in multiple components of the tissue (e.g. cell-types). For example, in the simplest scenario where patient tissue samples of healthy and disease patients are analyzed, these methods allow for the (un)supervised detection of a multicellular program describing the two populations of tissue samples. The multicellular programs inferred from these methods are interpretable, where the variability of a single feature (eg. gene) can be part of an exclusive and/or shared molecular process, similarly as novel cluster free methods for differential gene expression testing (106). Imagine the contrast of two cardiac tissues, one under hypoxic conditions and one in homeostasis. One could expect that relative to the oxygen supply, cells in a tissue, regardless of their type, will have a shared response in the context of the hypoxic pathway. However, the consequences of the activation of the hypoxic pathway in each cell-type may trigger specific molecular responses (107). Multicellular programs would capture both of these responses.

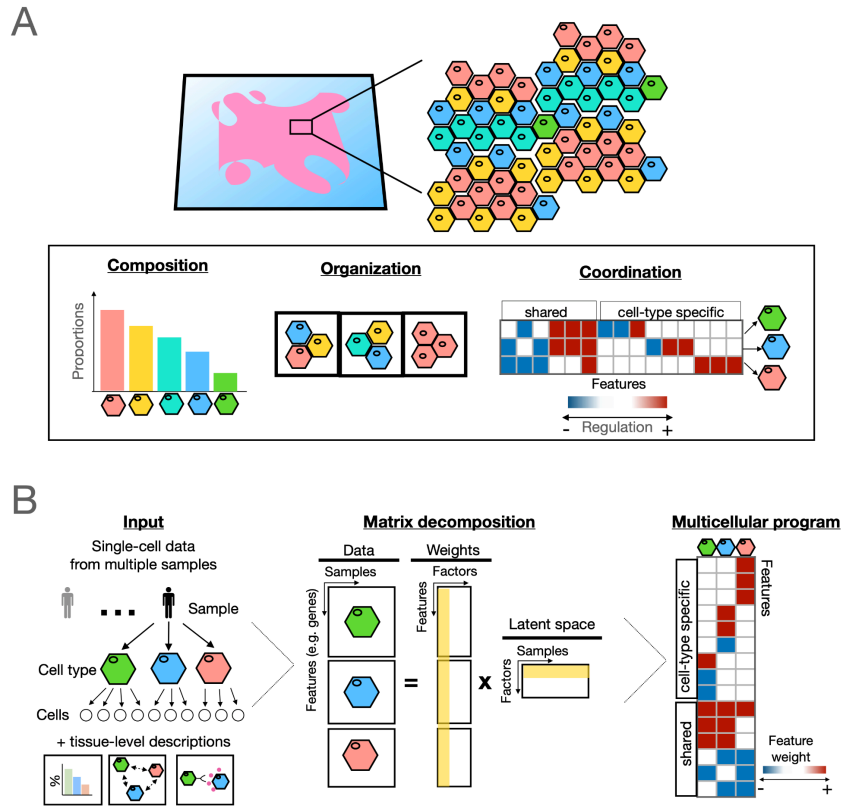

**Figure 3 Representations of the tissue state.**

A) With current single-cell technologies, descriptions of the tissue state (i.e. the multicellular function of the tissue) has focused on three major characteristics: 1) composition, the abundance of the tissue's components, 2) organization, that captures spatial or communication dependencies between the cells, and 3) coordination, that infers multicellular programs, representing co-varying molecular features (e.g. gene expression) in multiple components of the tissue.

B) Novel multicellular integration methods use available single cell atlases with multiple samples to infer multicellular programs. These methods typically start by converting single cell data across different conditions into a multi-view representation. Each view captures the aggregated gene expression profiles for cells of the same type within each sample. For the same reason, these methods rely on available cell taxonomies. Some methods can include additional compositional and organizational descriptions of the tissues analyzed. To infer multicellular programs these methods repurpose various classes of matrix decomposition approaches.

To infer multicellular information networks we propose a top-down approach that uses as a starting point multicellular programs inferred from single cell transcriptomics data (Figure 4A). We assume that multicellular programs are consequences of unmeasured tissue processes that can be represented as multicellular networks. Consider the example of multicellular programs inferred from tissues under hypoxic and homeostatic conditions. These programs encode distinct coordinated gene expression patterns across cell types or tissue niches, encapsulating the overall differences between the two conditions.

To reconstruct the topology of the information network of a multicellular program we propose to use organizational descriptions of tissues and predictive models (Figure 4B). This task is akin to the efforts in reconstructing gene regulatory networks from gene expression data (108). For the inference of gene regulatory networks, the presence of transcription factor motifs in cis-regulatory elements is used to constrain the relationships between regulators and genes. Similarly, for the inference of multicellular information networks, it is possible to use the spatial proximity or the communication potential of cell groups, together with their abundance, to limit putative functional links between them. A functional link between two cell groups could be inferred from a predictive model where the gene expression of a cell group is explained by the gene expression of the other. For example, one possibility could be to use cell communication prior knowledge to infer the effects of one cell group to the rest using network propagation as in (109). Interpretable machine learning models that incorporate multicellular communication beyond cell pairs and output relationships between extracellular and intracellular features (eg. gene-gene relationships) would be necessary. Finally, the multicellular information network inferred from this framework could be used for topological analysis or as a scaffold for dynamic modeling following the modeling principles of intracellular signaling (110) and gene regulatory networks (111) (Figure 4).

Inference of multicellular information networks following some of the principles discussed above have appeared in the literature primarily from the angle of cell-cell communication inference methods (112–116). However, a missing component in these approaches is the explicit use of multicellular coordination or tissue composition to constrain the inference of the cell-cell networks. Inclusion of the changes in tissue composition into the generation of multicellular information networks could link molecular coordination with structural changes in the tissue and initial implementations of this idea have appeared in the literature (117).

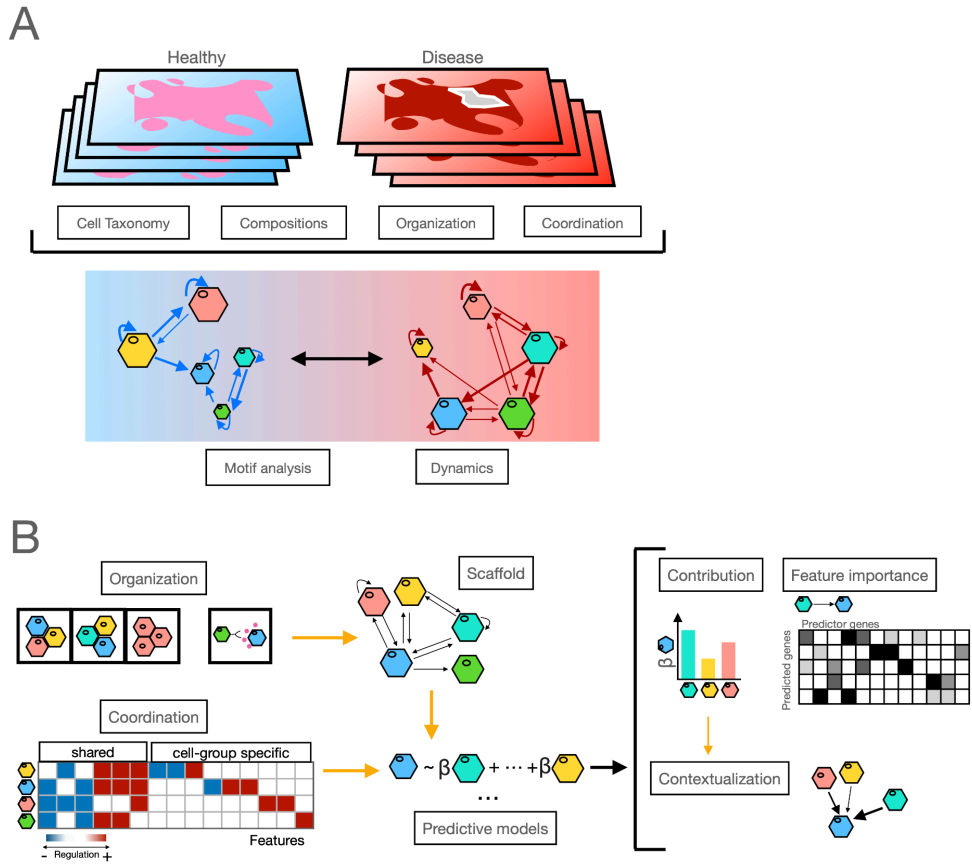

**Figure 4 Inference of multicellular information networks of a tissue function**

A) Multicellular coordination associated with a physiopathological process of interest could be represented in multicellular information networks. By relying on a cell taxonomy and integrating the changes in composition, organization, and coordination from cross-condition single cell data, the basic topology of a reinforced multicellular process can be inferred from single cell data. The networks then can be used for topological and dynamical modeling as it has done with intracellular processes.

B) To infer a multicellular network, we propose to build a scaffold of cell dependencies using organizational representations of tissues that use spatial data or communication inference tools. In parallel, we propose the use of multicellular programs to constrain predictive models that aim to explain the expression of the population of a target cell group with the information of the rest of the groups. These interpretable predictive models on the one hand can help to infer interaction weights in the multicellular information network, and on the other can infer gene-gene relationships. An iterative process of contextualization for each cell group will generate the putative multicellular network.

Our top-down approach assumes that multicellular programs that differentiate healthy and disease tissues capture reinforced long-term coordinated processes of groups of cells. Even though the measurements that are used to infer multicellular information networks come from independent locations and patients, there is evidence from the analysis of cross-condition bulk and single cell transcriptomics data that disease processes of the same type show a degree of convergence. For more than 20 years, biomedical researchers have contrasted the bulk gene expression of patient's tissues from distinct clinical conditions (118), and at the population level they have observed a convergence of gene expression patterns associated with disease processes, generating disease signatures (119). Similarly, recurrent changes in compositional and spatial organizations in tissues are observed when similar contrasts are performed from dissociated or spatial single-cell data (43, 60, 120). We propose that these consistent molecular profiles in tissues, indicative of a shared regulatory disease process, are the results of reinforced multicellular events originating from diverse cell populations. Hence, these recurrent molecular patterns could serve as a foundation for inferring shared multicellular information networks in disease contexts.

**Utility of studying multicellularity**

Complementing cell taxonomies with the study of multicellular coordination could help us to better understand tissue function, since it will integrate the descriptions of the composition and organization of tissues with the functional dependencies between the diverse cell groups that compose them. We believe that this multicellular functional perspective of single cell data could promote the emergence of novel analysis strategies that could synergize the knowledge of the fields of mathematical physiology, ecology, and systems biology. While we acknowledge the necessity to harmonize the single-cell data generated from large community efforts, we also believe that the study of multicellular coordination with current data is feasible. Moreover, these multicellular descriptions of tissues rather than individual cells provide a more functional description, that thus is likely to better capture commonalities of medical relevance across patient cohorts.

As previously mentioned, studying coordinated multicellular processes enables the assignment of functional roles to the diverse groups of cells identified in single cell atlases. For example, groups of cells that act as first-responders or communication-hubs in disease processes could be identified by contrasting the number and strength of interactions in multicellular information networks across tissue states. Nevertheless, to study the state of tissues and their functions we require to develop a new vocabulary. How can we describe or call multicellular interactions? Signaling pathway activities in single cells become blurry as a tissue-level higher picture is described from the multicellular perspective, but physiological descriptions such as inflammation, healing, or immune responses are too broad or unspecific. One possibility is to describe cells in terms of their relationship with others, reflecting known ecological terms such as competition, mutualism, etc. (121). We agree with this view that understanding the organization of the information multicellular network in terms of minimal functional modules, together with their rules, could provide a new paradigm to study tissues from the molecular perspective.

Moreover, a major challenge in the analysis of single-cell and spatial data is the cross-condition comparison of whole tissue samples (93). To solve this question, some methods define distances between whole tissue samples based on the combination of the composition of cell groups together with their molecular differences (122–125). However, the interpretability of these strategies is limited, since it is hard to quantify which cell group contributes the most in the variability of tissue samples. Multicellular programs are interpretable data embeddings that incorporate the variability between distinct tissue samples across multiple cell groups or niches, and thus can be used as a methodological alternative for (un)supervised cross-condition comparisons (26–30). Moreover, extensions of these methods already allow the incorporation of compositional and organizational features of tissues in the estimation of multicellular programs and the quantification of their importance in sample variability (28). The opportunity to quantify the contribution of the compositional, organizational and multicellular information in the description of disease processes could aid in the definition of patient groups.

Furthermore, a multicellular perspective opens new avenues for the meta-analysis of distinct patient cohorts, disease processes, and technologies. The inference of multicellular information networks provides a complementary strategy to integrate single-cell and spatial data that goes beyond cell-type calling and mapping. Moreover, multicellular representations can be highly beneficial for the biomedicine community, since they could help to understand if chronic organ failures that are driven by distinct initial causes converge to similar multicellular information networks, limiting the search space of therapies. Finally, multicellular descriptions of tissues represent a bridge to other scales such as bulk omics, that have profiled larger patient cohorts in comparison to their single cell counterparts. For example, this novel multi-scale integration could help in the rethinking of bulk deconvolution strategies that have mainly been based on the understanding of tissues from the taxonomic perspective (126). If molecular profiles both at the single-cell and bulk level are understood as footprints of unmeasured multicellular processes, an alternative perspective would be to deconvolute the variability of gene expression from bulk profiles in terms of the compositional and multicellular description of tissues (28).

Ultimately, we believe that the combination of (i) ongoing single-cell integration efforts, (ii) rich cell taxonomies, (iii) novel multicellular integration computational methods, and (iv) public bulk -omics datasets will allow us to meta-analyze human disease processes from a multicellular tissue-centric perspective and to understand multicellular coordination.

**Conclusions**

To properly understand tissue function from a molecular perspective it is fundamental to study the regulatory mechanisms that enable cells to couple their intra- and extracellular processes to work as a community, despite the expected diversity of responses and dynamics of single cells. Here, we discussed the possibilities of describing at a high-level the topology and dynamics of multicellular information networks in tissues by integrating compositional, organizational and functional descriptions of single-cell data from dissociated or intact tissues. We argued that a multicellular perspective would help in the assignment of functional roles of groups of cells in physiological processes and in the integration of data across patient cohorts and technologies.

To prove the potential rules that govern multicellular information networks, we need to combine novel methodologies that enable to study multicellular systems at the molecular and single cell level under various controlled conditions. As multicellular systems emulating tissues continue improving (127), there is an opportunity to take advantage of the resolution of single-cell profiling (128, 129), live imaging and emerging live molecular profiling (130) to measure the effects that multiple perturbations have on synthetic tissues and confirm the topologies of multicellular information networks.

We believe that while tissue taxonomy is necessary, a parallel effort focused on the description of tissue states as communities will allow us to contextualize the observed diversity of cells reported in single cell atlasing efforts. Human cells do not work in isolation and understanding disease as a disruption of the sociobiology of cells could ultimately impact the way we develop novel therapeutics.

**Acknowledgements**

We acknowledge the support of German Science foundation (DFG) through the CRC 1550 "Molecular Circuits of Heart Disease", SPP 2395 "Local and Peripheral Drivers of Microglial Diversity and Function", and the Federal Ministry of Education and Research (BMBF) through the funding 01EJ2201B of the project "Curefib". We thank Daniel Dimitrov, Pau Badia i Mompel, Jan Lanzer, Jovan Tanevski, and Miguel Ibarra for discussions that shaped the manuscript.

**Conflict of interests**

JSR reports funding from GSK, Pfizer and Sanofi and fees/honoraria from Travere Therapeutics, Stadapharm, Astex, Pfizer and Grunenthal.